\documentclass[aps,twocolumn,superscriptaddress,longbibliography]{revtex4-1}

\usepackage{graphicx}% Include figure files
\usepackage{CJK}
\usepackage{dcolumn}
\usepackage{bm}% bold math
\usepackage[colorlinks=true,bookmarks=false,citecolor=blue,linkcolor=blue,hyperfootnotes=true,urlcolor=blue]{hyperref}
\usepackage{color}
\usepackage{comment}
\usepackage{sidecap}
\usepackage{nicefrac}
\usepackage[acronym]{glossaries}

\newacronym{CDW}{CDW}{charge density wave}
\newacronym{HTT}{HTT}{high temperature tetragonal}
\newacronym{LTT}{LTT}{low temperature tetragonal}
\newacronym{LTO}{LTO}{low temperature orthorhombic}
\newacronym{SEM}{SEM}{scanning electron microscopy}
\newacronym{TEM}{TEM}{transmission electron microscopy}
\newacronym{2D}{2D}{two dimensional}

\begin{document}

\title{Persistent Charge Density Wave Memory in a Cuprate Superconductor}

\author{X. M. Chen}
\email{xmchen@lbl.gov}
\altaffiliation[Present address: ]{Advanced Light Source, Lawrence Berkeley National Laboratory, Berkeley, CA 94720, USA}
\affiliation{Condensed Matter Physics and Materials Science Department, Brookhaven National Laboratory, Upton, New York 11973, USA}

\author{C. Mazzoli}
\affiliation{National Synchrotron Light Source II, Brookhaven National Laboratory, Upton, New York 11973, USA}

\author{Y. Cao}
\affiliation{Condensed Matter Physics and Materials Science Department, Brookhaven National Laboratory, Upton, New York 11973, USA}

\author{V. Thampy}
\altaffiliation[Present address: ]{Stanford Synchrotron Radiation Lightsource, SLAC National Accelerator Laboratory, CA 94025, USA}
\affiliation{Condensed Matter Physics and Materials Science Department, Brookhaven National Laboratory, Upton, New York 11973, USA}

\author{A. M. Barbour}
\author{W. Hu}
\affiliation{National Synchrotron Light Source II, Brookhaven National Laboratory, Upton, New York 11973, USA}

\author{M. Lu}
\affiliation{Center for Functional Nanomaterials, Brookhaven National Laboratory, Upton, New York 11973, USA}

\author{T. Assefa}
\author{H. Miao}
\author{G. Fabbris}
\author{G. D. Gu}
\author{J. M. Tranquada}
\author{M. P. M. Dean}
\email{mdean@bnl.gov}
\affiliation{Condensed Matter Physics and Materials Science Department, Brookhaven National Laboratory, Upton, New York 11973, USA}
\author{S. B. Wilkins}
\email{swilkins@bnl.gov }
\affiliation{National Synchrotron Light Source II, Brookhaven National Laboratory, Upton, New York 11973, USA}
\author{I. K. Robinson}
\email{irobinson@bnl.gov}
\affiliation{Condensed Matter Physics and Materials Science Department, Brookhaven National Laboratory, Upton, New York 11973, USA}
\affiliation{London Centre for Nanotechnology, University College, Gower St., London, WC1E 6BT, UK}

% user macros
\def\mathbi#1{\ensuremath{\textbf{\em #1}}}
\def\Q{\ensuremath{\mathbi{Q}}}
\def\q{\ensuremath{\mathbi{q}}}

\def\LBCO{La$_{1.875}$Ba$_{0.125}$CuO$_4$}
\def\LSCO{La$_{1.875}$Sr$_{0.125}$CuO$_4$}

\def\LBCOx{La$_{2-x}$Ba$_x$CuO$_4$}
\def\LSCOx{La$_{2-x}$Sr$_x$CuO$_4$}
\def\LNSCOx{La$_{2-x-y}$Nd$_y$Sr$_x$CuO$_4$}

\newcommand{\microns}{$\mathrm{\mu}$m}
\newcommand{\angstrom}{\mbox{\normalfont\AA}}
\date{\today}

\begin{abstract}
Although \gls*{CDW} correlations appear to be a ubiquitous feature of the superconducting cuprates, their disparate properties suggest a crucial role for coupling or pinning of the \gls*{CDW} to lattice deformations and disorder.  While diffraction intensities can demonstrate the occurrence of \gls*{CDW} domain formation, the lack of scattering phase information has limited our understanding of this process. Here, we report coherent resonant x-ray speckle correlation analysis, which directly determines the reproducibility of \gls*{CDW} domain patterns in \LBCO{} (LBCO~1/8) with thermal cycling. While CDW order is only observed below 54~K, where a structural phase transition results in equivalent Cu-O bonds, we discover remarkably reproducible CDW domain memory upon repeated cycling to temperatures well above that transition. That memory is only lost on cycling across the transition at 240(3) K that restores the four-fold symmetry of the copper-oxide planes.  We infer that the structural-domain twinning pattern that develops below 240~K determines the \gls*{CDW} pinning landscape below 54~K. These results open a new view into the complex coupling between charge and lattice degrees of freedom in superconducting cuprates.

\end{abstract}

%\todo[inline]{Abstract should stress lack of exp. probes?}

%\keywords{}
\maketitle

Holes doped into the Mott insulating parent compounds of the high temperature superconducting cuprates experience strong interactions with the antiferromagnetic background and with each other, as well as with the lattice in which they reside \cite{Keimer2015}. The Hubbard Hamiltonian, often used to model the cuprates, predicts that \gls*{CDW} fluctuations are an intrinsic property of strongly interacting electrons in pristine, undistorted 2D square lattices \cite{Zheng2017stripe, Huang2017numerical}. \gls*{CDW}s have indeed been observed in essentially all hole doped cuprates, but with distinct transition temperatures, correlation lengths and wavevectors \cite{Tranquada1995, Ghiringhelli2012, Comin2014, daSilvaNeto2014, Thampy2014, tabis2014, Miao2017high, miao2018incommensurate}. This is epitomized by comparing two very similar cuprates that have slightly different low temperature crystal structures: \LSCO{} (LSCO~1/8) and LBCO~1/8. LSCO~1/8 has weak \gls*{CDW} correlations and is a bulk superconductor below its transition temperature of $T_c=28$~K \cite{Thampy2014, Croft2014, Christensen2014, Wu2012, LSCO_note}; whereas LBCO~1/8 has well correlated \gls*{CDW} order that almost completely suppresses bulk superconductivity into what is proposed to be a \gls*{2D} superconducting pair density wave state \cite{Moodenbaugh1988, Li2007, Berg2007, Fradkin2015, Xu2014}. As the main difference between these compounds is a subtle change in the crystal structure \cite{Axe1989}, it is evident the lattice that hosts the \gls*{CDW} correlations has a dramatic influence on the properties of the \gls*{CDW} and the superconducting ground state. 

Multiple types of lattice deformation or disorder are present in cuprates including that of interstitial oxygen atoms, chemical substitutions as well as local and long-range tilting of the Cu-O octahedra, all of which have been proposed as possible \gls*{CDW} pinning features that stabilize CDW order \cite{Tranquada1995, Bozin1999charge, Kivelson2003, Hanaguri2004checkerboard, Vojta2009, Nie2014quenched, campi2015inhomogeneity, Wu2015}. One widely discussed lattice deformation in this context is the \gls*{LTT} phase in LBCO, in which octahedral tilts define a preferential direction for stripe pinning within each CuO$_2$ plane \cite{Tranquada1995,Kivelson2003,Axe1989}. While the \gls*{LTO} phase is common for both LBCO and LSCO, the \gls*{LTT} structural transition is a special feature in LBCO that coincides with the CDW formation. Fully understanding this process is, however, hampered by a lack of  experimental techniques that are sensitive to the phase of the CDW order parameter.  To this end, we implemented coherent resonant x-ray speckle correlation analysis as a tool for probing \gls*{CDW} domain pinning in the cuprates, choosing LBCO~1/8 as the model system due to its \gls*{LTT} structure and particularly large \gls*{CDW} correlation length. We discover strikingly reproducible CDW domain formation upon repeated thermal cycling well above its transition temperature and show that the CDW pinning memory is defined by structures that form at the \gls*{LTO} transition rather than disorder or the \gls*{LTT} transition that appears alongside \gls*{CDW} order.

\begin{figure}
    \includegraphics[width=0.35\textwidth]{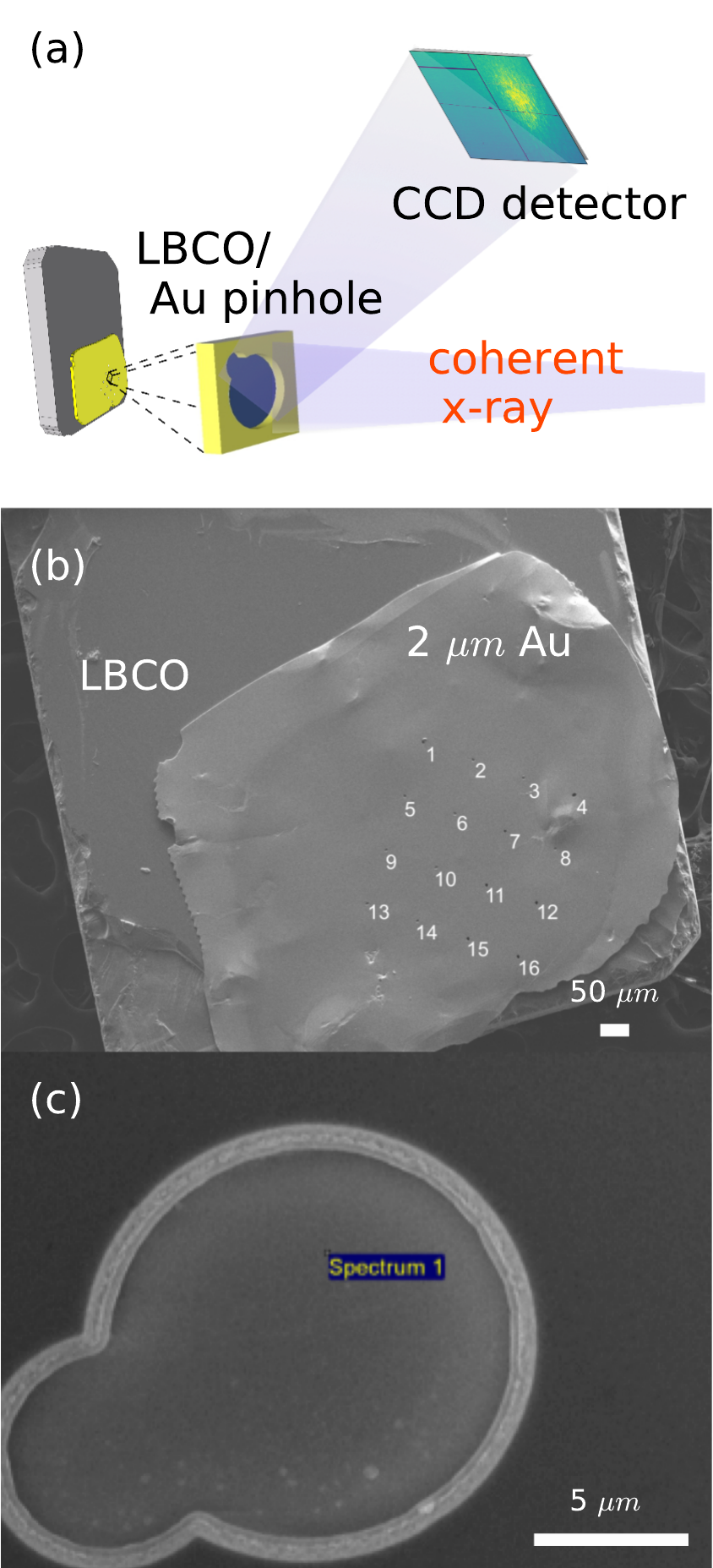}
    \caption{Experimental setup. (a) The scattering geometry in which coherent x-rays illuminate the masked LBCO~1/8 sample and the \gls*{CDW} Bragg peak is measured on a CCD detector. (b) A \gls*{SEM} image of the LBCO~1/8 single crystal with a Au mask fixed on top. Numbers 1-16 indicate the sixteen pinholes drilled in the Au mask using a focused ion beam (FIB) prior to being fixed on the crystal. (c) A zoomed \gls*{SEM} image of pinhole \#10, through which most of our data were taken.}
    \label{fig1}
\end{figure}

\section*{Experiment and results}
Although \gls*{CDW} pinning can dramatically change the superconducting ground state, directly observing \gls*{CDW} pinning and tracking its changes with temperature are very challenging tasks.  Much of the difficulty is that one must combine a technique that is sensitive enough to detect the \gls*{CDW} domain spatial arrangement, which we will call ``texture", with the ability to reproducibly illuminate the same sample volume over a wide temperature range. Figure~\ref{fig1} shows how we have addressed the problem using coherent resonant x-ray diffraction: a scattering method that measures the interference between scattering from different domains in the \gls*{CDW} texture as a ``speckle" pattern. Our innovation, reported here, is to attach a mask with a microscopic pinhole to the sample, which we overfill with the coherent x-ray beam to ensure we illuminate the same sample volume despite possible temperature induced drifts of the sample position.

An LBCO~1/8 single crystal was aligned to the \gls*{CDW} Bragg condition at wavevector $\Q{} = (0.236, 0, 1.5)$ (see Methods). We furthermore tuned the x-ray energy to the Cu $L_3$-edge resonance around 931~eV in order to enhance our sensitivity to the weak CDW. In this condition the incident x-ray angle is $\theta_i = 88^{\circ}$ with respect to the $[001]$ sample surface and the detector angle is $2\theta = 119^{\circ}$. Such an approach has been used extensively to study the average \gls*{CDW} properties \cite{Abbamonte2005, Wilkins2011, Ghiringhelli2012, Comin2014,daSilvaNeto2014, Thampy2014, Hashimoto2014, tabis2014}. The very high coherent flux ($10^{13}$ photons/s) at the 23-ID-1 beam line at the National Synchrotron Light Source II opens up the possibility of observing coherent interference between the domains of a weakly scattering order parameter such as cuprate CDWs \cite{Chen2016, Thampy2017static}. This was configured to produce a beam of approximately 20~\microns{} at the sample overfilling the 10~\microns{} pinhole in the Au mask attached to the sample (see Methods). 

\begin{figure}
    \includegraphics[width=0.5\textwidth]{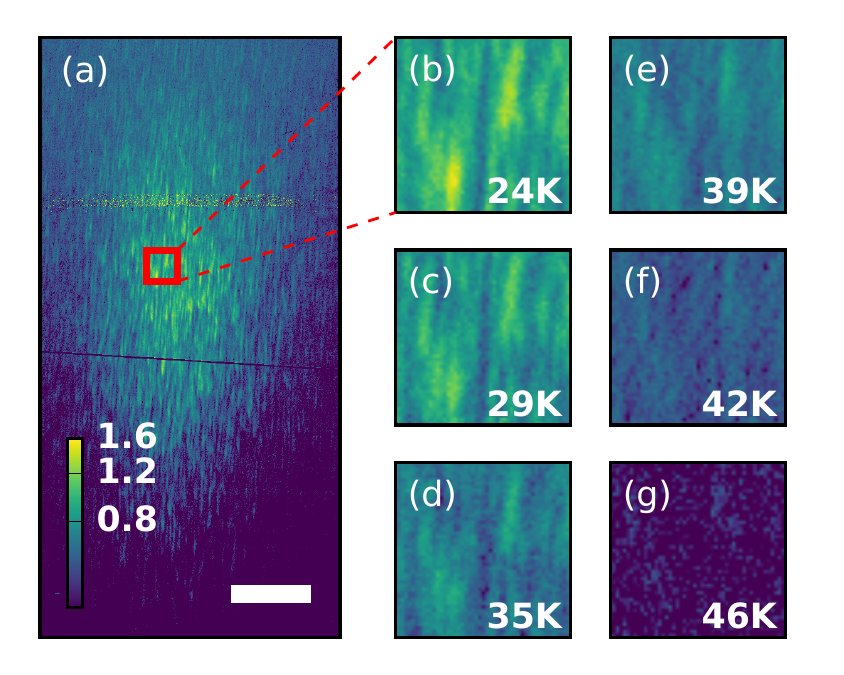}
    \caption{Temperature dependence of the \gls*{CDW} speckle positions within the ordered state. (a) A detector image of the LBCO~1/8 \gls*{CDW} Bragg peak at 24~K at $\Q = (0.236, 0, 1.5)$. The color-bar denotes intensity in photons/second, and the white bar at the bottom indicates 100 detector pixels (0.0025 r.l.u). (b-g) The temperature dependence of speckle positions as the temperature was raised from 24 to 46~K. Zoomed-in speckle images are taken from the same detector area indicated by the red box in (a). Despite the broadening and weakening of the \gls*{CDW} Bragg peak, the speckles tend to persist in similar locations to $46$~K, above which the speckle intensity becomes too low and noise dominates the signal.}
    \label{fig2}
\end{figure}

\begin{figure}
    \includegraphics[width=\linewidth]{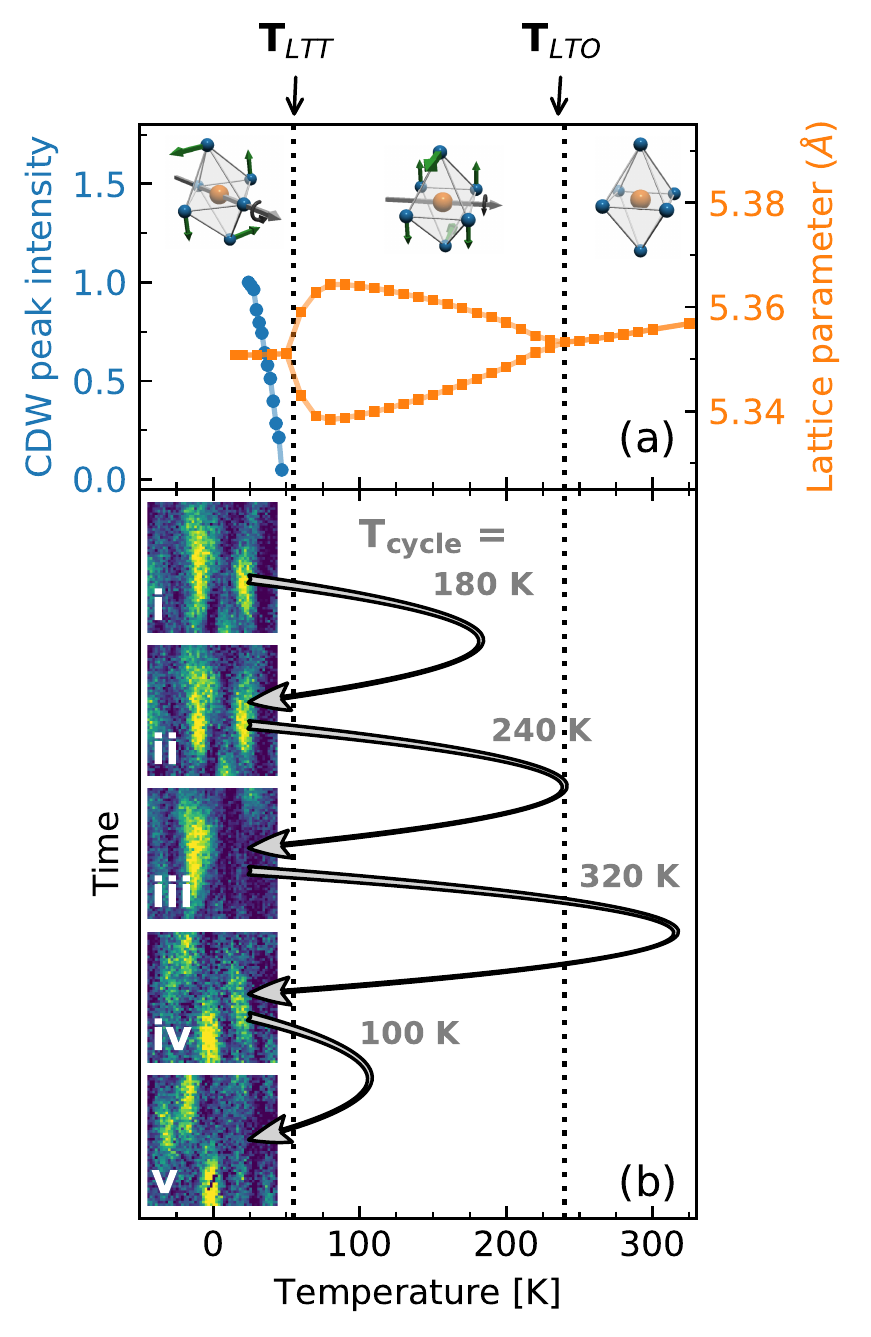}
    \caption{(a) Temperature dependence of normalized CDW peak intensity (blue) and lattice parameters reproduced from \cite{Bozin2015} (orange). The CDW transition coincides with the \gls*{LTT} structural transition at $T_{\text{LTT}}$. A second lattice transition occurs at $T_{\text{LTO}}$. The inset images represent the octahedral tilts associated with the \gls*{LTT}, \gls*{LTO} and \gls*{HTT} phases. (b) CDW speckle images taken at 24~K. The black arrows indicate the temperature cycling between images. $T_{\text{cycle}}$ indicates the highest temperature that sample was brought to during a temperature cycle.}
    \label{fig3}
\end{figure}

Figure \ref{fig2}(a) plots a detector image at the \gls*{CDW} Bragg condition, which is zoomed in panels (b)-(g). The extent of the observed peak (in this case $\sim 4\times10^{-3}$~\AA{}$^{-1}$) is inversely proportional to the \gls*{CDW} correlation length (or domain size) similar to what is seen in conventional scattering. The speckle modulations on top of the peak envelope arise from coherent interference between different \gls*{CDW} domains. The average size and elongated shape of the speckles on the detector is determined primarily by the geometry of the experiment and the extent of the illuminated sample volume, which is fixed by the 10~\microns{} pinhole and $\sim100$~nm x-ray penetration depth [Fig.~\ref{fig1}(a)] \cite{peakshape}. Therefore, the speckle size and shape do not immediately provide information on the sample properties. The speckle locations, however, are highly sensitive to the CDW domain positions and can therefore be used to test for changes in the \gls*{CDW} domain texture as a function of temperature \cite{Pierce2003, Chesnel2012, Chesnel2016shaping}. In LBCO~1/8, static \gls*{CDW} order exists at low temperatures with a correlation length of $240$~\AA{}. This can be directly verified with x-rays by noting that the speckle pattern does not vary as a function of time as reported previously \cite{Chen2016, Thampy2017static}.  With increasing temperature the \gls*{CDW} correlation length decreases and the CDW Bragg peak disappears coincident with $T_{\text{LTT}} = 54$~K (Fig.~S3, \cite{Hucker2011,Abbamonte2005,Wilkins2011,Hucker2013,DeanLBCO2013, Chen2016}). The temperature dependence of the CDW peak intensity is explicitly compared with the \gls*{LTT} to \gls*{LTO} structural phase transition in Fig.\ref{fig3}~(a). We measured a range of temperatures from $24$ up to $46$~K (the highest temperature for which we have sufficient \gls*{CDW} signal in order to observe the speckle locations). At all temperatures, the speckle locations were found to persist as illustrated in Fig.~\ref{fig2}(b)-(g). 

Next we studied the reproducibility of the \gls*{CDW} domain texture after thermal cycling. Speckle images were taken at $24$~K successively before and after the sample temperature was cycled to $T_{\text{cycle}}$ as we illustrate in Fig.~\ref{fig3} (b). After cycling to $T_{\text{cycle}}=180$~K, a temperature well above the \gls*{CDW} transition temperature, the speckle patterns are strikingly similar [see panels (i) and (ii)]. However, as the sample was heated to higher temperatures, the degree of reproducibility dropped, and by $T_{\text{cycle}} = 320$~K, their positions completely changed as seen in panels (ii) to (iv). 

In order to quantitatively compare speckle positions, and identify the onset temperature of this change, we calculated the normalized cross-correlation function. Speckle images were background subtracted as described in the Supplementary Information Section S1. These images are then represented as $M \times N$ matrices $A_{m,n}$ and $B_{m,n}$ where $m$ and $n$ are row-column indices. Cross-correlations matrices are calculated via
\begin{equation}
    A_{m,n} \ast B_{m,n} = \sum\limits_{m' = -M}^{M}\sum\limits_{n' = -N}^{N} A_{m',n'} B_{m+m', n+n'}.
    \label{eq:cross-correlation}
\end{equation}
Here, $A_{m,n}$ and $B_{m,n}$ were taken to be  $M = 200 \times N = 200$-pixel images under the \gls*{CDW} peak measured before and after thermally cycling the sample, which include $\sim 30-50$ speckles. When $A_{m,n}$ and $B_{m,n}$ have the same or similar speckle patterns, the correlated intensity features a peak that we sum over to obtain a single normalized speckle cross-correlation coefficient \cite{Pierce2003, Chesnel2012, Chesnel2016shaping}
\begin{equation}
    \xi = \frac{\sum\limits_{\text{speckle}}A_{m,n}\ast B_{m,n}}{\left(\sum\limits_{\text{speckle}}A_{m,n}\ast A_{m,n} \sum\limits_{\text{speckle}}B_{m,n}\ast B_{m,n}\right)^{1/2}}.
    \label{eq:norm_cross-correlation}
\end{equation}

\begin{figure}
    \includegraphics[width=\linewidth]{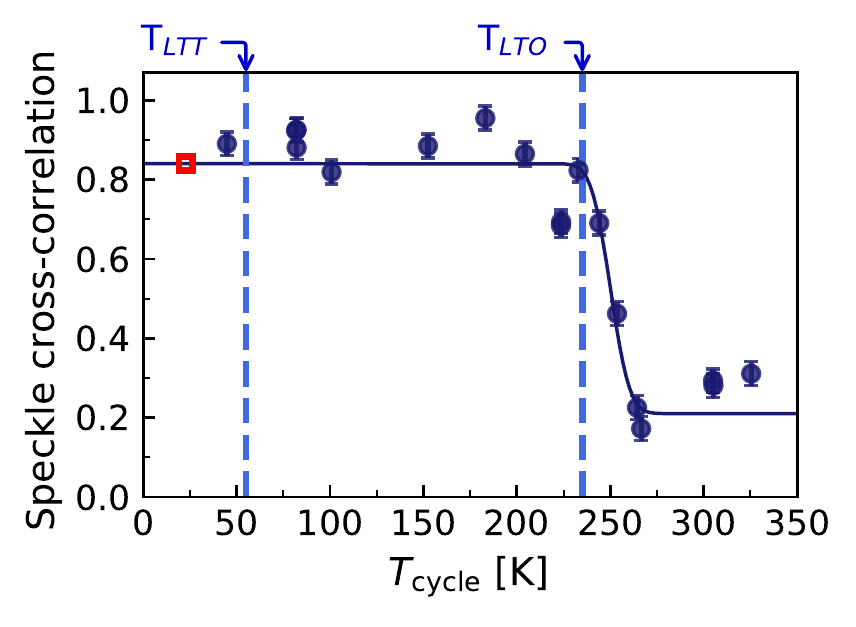}
    \caption{Normalized speckle cross-correlation coefficient, $\xi$, of data before and after temperature cycling to different values of $T_{\text{cycle}}$. Images of $200 \times 200$ detector pixels were used for this calculation. An error function fit gives a de-correlation onset temperature of $240(3)$~K, which coincides with the LTO structural phase transition. The red box indicates 24~K the temperature where all the speckle images were sampled for this figure. Error bars are estimated by taking the standard deviation of repeated, nominally equivalent, measurements.
    } \label{fig4}
\end{figure}

In this equation, $\sum\limits_{\text{speckle}}$ denotes summing over the peak in the cross-correlation matrix, corresponding to just over one speckle ($\sim 4 \times 16$ pixels) in size. The value is normalized by dividing by the auto-correlations of $A_{m,n}$ and $B_{m,n}$, such that two identical images have $\xi=1$. In Fig.~\ref{fig4}, we show $\xi$ for various cycle temperatures showing a transition from high to low speckle reproducibility well above the \gls*{CDW} ordering temperature. We fit $\xi$ versus $T_{\text{cycle}}$ using an error function, which provides a reasonable phenomenological description of the shape of the transition. The onset temperature of de-correlation was $240(3)$~K, which coincides with the LTO structural phase transition at $T_{\text{LTO}} = 236(5)$~K \cite{Hucker2011}. This transition involves rotations of the Cu-O octahedra around the $\langle 110 \rangle _{\text{HTT}}$ direction and since nearest-neighbor octahedra rotate in opposite directions, the unit cell volume increases by a factor of two, with $a_o \approx b_o \approx \sqrt{2}a$. Below this threshold the majority of the speckles reproduce with $\xi=0.84(9)$; above the transition $\xi=0.21(4)$, which corresponds to the value expected in a random domain distribution as estimated by computing $\xi$ after flipping one of the images about a horizontal axis. We further confirmed that different regions within the speckle pattern shown in Fig.~\ref{fig2}(a) and data taken through another pinhole reproduce the same phenomenology within error bars.

\section*{Discussion}
This work establishes that the \gls*{CDW} domains are strikingly reproducible upon temperature cycling well above the \gls*{LTT} and \gls*{CDW} transitions, up to $T_{\text{cycle}} = 240(3)$~K. Disorder and structural distortions are present in almost all cuprates and have been discussed extensively in connection with \gls*{CDW} pinning in the cuprates \cite{Alloul2009, Tranquada1995, Bozin1999charge, Kivelson2003, Hanaguri2004checkerboard, Vojta2009, Nie2014quenched, campi2015inhomogeneity, Wu2015}. In \LBCOx{} the most prominent form of disorder is La/Ba cation substitution and the most obvious structural features come from the \gls*{LTO} and \gls*{LTT} phase domain boundaries. 

Random La/Ba distribution, or other forms of chemical disorder, appear to play a role in pinning the \gls*{CDW} into a static ground state and in setting an upper limit on the \gls*{CDW} correlation length \cite{Alloul2009, Vojta2009, campi2015inhomogeneity, Nie2014quenched, Chen2016}. Chemical disorder such as this will, however, always be fixed at all temperature considered in this study and therefore cannot explain the sharp transition in the observed memory effect upon cycling above 240~K.

In the context of structural \gls*{CDW} pinning, the importance of the \gls*{LTT} phase has been emphasized since the discovery of \gls*{CDW} order in the cuprates \cite{Tranquada1995}. This phase features rows of oxygen atoms along the $\langle 100 \rangle _{\text{HTT}}$ directions displaced in/out of the CuO$_2$ planes, which appear particularly suitable for pinning \cite{Axe1989, Tranquada1995, Hucker2012structural}. The concomitance of the \gls*{CDW} and \gls*{LTT} transitions, as well as the particularly long \gls*{CDW} correlation length in LBCO and La$_{1.6-x}$Nd$_{0.4}$Sr$_x$CuO$_4$ relative to other cuprates such as LSCO, is surely due to coupling between the \gls*{LTT} and \gls*{CDW} phases \cite{Tranquada1995, Fujita2004, Wilkins2011, Hucker2013, DeanLBCO2013, Thampy2014, Baity2016collective, Miao2017high}. It is, consequently, highly intriguing that the \gls*{CDW} memory persists well above this transition. Figure~\ref{fig4} shows a sharp cross-over from near-complete to minimal reproduciblity at 240(3)~K. This threshold temperature coincides with the \gls*{LTO} phase transition at 236(5)~K \cite{Hucker2011, Bozin2015}, implying that the pinning configuration is effectively determined at this phase transition, where octahedral tilts break the four-fold symmetry of copper-oxide planes. At this \gls*{LTO} transition \gls*{LTO} twin domain boundaries form. Based on strain considerations, these are expected to occur along the $\langle 100 \rangle _{\text{HTT}}$ lattice directions, as confirmed by \gls*{TEM} imaging \cite{Zhu1994, Chen1991low, Chen1993}.  Early structural studies of LBCO demonstrated that the \gls*{LTT} and \gls*{LTO} tilt patterns can be conceptualized as superpositions of one-another \cite{Axe1989, Hucker2012structural} and superposition-based arguments naturally justified the observation that \gls*{LTT}-like tilts occur at \gls*{LTO} domain boundaries in the cuprates  \cite{Zhu1994, Chen1991low, Chen1993}. This interaction opens the possibility that the \gls*{LTT} domain boundary configuration can be inherited from the \gls*{LTO} domain boundary configuration. This would explain how the \gls*{CDW} onset temperature coincides with the \gls*{LTT} transition, but the memory effect matches the  \gls*{LTO} transition.

Another significant observation is that, within the \gls*{CDW} phase,  the speckle locations do not change with temperature (see Fig.~\ref{fig2}), despite the fact that the \gls*{CDW} correlation length varies with temperature \cite{Fujita2004,Hucker2011,Hucker2013,Wilkins2011,Miao2017high}. This can be explained by uniform expansion of \gls*{CDW} domains around their pinning centers on cooling, as speckle locations are primarily dependent on the locations of the domains and largely insensitive to their size (see Supplementary Information Section~S3).

Almost all cuprates show some form of orthorhombic structural symmetry breaking, so the CDW memory effects discovered here may well be shared by many different cuprate species. Testing whether other cuprates, including those without a \gls*{LTT} phase, show similar behavior will be important in future work. We further emphasize that the experimental configuration presented here, in which sample drift problems are avoided by attaching a mask directly to the sample, is also applicable to applied current or laser excitation as well as to the magnetic field dependent behavior studied previously in magnetic alloys \cite{Pierce2003, Chesnel2012, Chesnel2016shaping}. This complements what can be achieved with scanning tunneling spectroscopy, which has superior spatial resolution, but is limited to cleaved surfaces \cite{Hanaguri2004checkerboard}. The use of coherence and resonant x-ray scattering further opens the possibility to study charge, spin and orbital order parameters in other quantum materials with better resolution than is possible with current microdiffraction techniques \cite{campi2015inhomogeneity}.

In conclusion, we present the first application of coherent resonant soft x-ray speckle correlation analysis to study the \gls*{CDW} domain hysteresis in LBCO~1/8. We uncover remarkably reproducible CDW domain formation upon temperature cycling far above the 54~K \gls*{CDW} transition, before the CDW pattern is almost completely reconfigured by cycling above  240(3)~K. The \gls*{CDW} order, which is associated with the dramatic suppression of superconductivity in this sample, experiences a pinning landscape that is determined by structural domains that form at the \gls*{LTO} phase transition. Our results open a new route to studying the complex interplay between lattice and charge degrees of freedom in quantum materials at cryogenic temperatures.

\appendix

\section*{Methods}

\subsection*{Sample preparation \label{sample}}
Single crystals of LBCO~1/8 were grown using the floating zone method, and characterized extensively in previous studies \cite{Wilkins2011, Thampy2013, DeanLBCO2013, Miao2017high, miao2018incommensurate}, all indicating excellent sample quality. A $2~\mathrm{\mu m}$ gold (Au) film was evaporated onto a free-standing 200~nm Si$_3$N$_4$ membrane. Arrays of $10~\mathrm{\mu m}$ asymmetric pinholes were drilled into the film using a focused ion beam yielding the shape shown in  Fig.~\ref{fig1} (c) \cite{bCDI}.  This assembly was pressed onto the LBCO single crystal with Poly Methyl Methacrylate (PMMA) as a glue. The excess PMMA that filled in the pinholes was then removed using reactive ion etching. The resulting sample was imaged in a JEOL 7600F scanning electron microscope [Fig.~\ref{fig1}, (b) and (c)]. In this paper we index the crystal using the  \gls*{HTT}  unit cell where $a = b = 3.78$ and $c =13.28$~\AA{}. Correlation length here is defined as $a/\text{HWHM}$ where HWHM is the half width at half maximum of the peak in r.l.u.

\subsection*{X-ray scattering and Imaging \label{scattering}}
Data were taken at the 23-ID-1 beamline at the National Synchrotron Light Source II (NSLS II) which is optimized to deliver high coherent flux at the sample. The beam was energy-dispersed in the vertical plane with a grating and focused onto a $20~\mathrm{\mu m}$ pinhole approximately 5~mm from the sample to which the Au mask with a $10~\mathrm{\mu m}$ pinhole was attached. In this configuration, the transverse coherence of the beam is greater than the extent of the illuminated region of the sample and the longitudinal coherence of the beam is of order $2~\mathrm{\mu m}$. The X-ray energy was further tuned to the Cu $L_3$ absorption edge around 931~eV to enhance the signal from the \gls*{CDW} peak. The positions of the pinholes were determined by scanning the sample through the X-ray beam using the LBCO Cu fluorescent yield. Data were collected using a fast CCD \cite{fccd_camera} with a $30 \times 30$ \microns{}$^2$ pixel size placed 340~mm from the sample. Images were read out every 5~s for a total collection time of approximately 10~min. 

\bibliographystyle{naturemag}
\bibliography{refs}

\begin{thebibliography}{10}
\bibitem{bCDI}
\bibinfo{note}{Although any stable shape is adequate for speckle correlation
  analysis, we chose to make a two-lobe structure. This decision was taken to
  break circular symmetry, which should improve the prospects for obtaining
  real-space images via Bragg Coherent Diffractive Imaging in the future. Such
  attempts have thus far been unsuccessful.}
\end{thebibliography}

\begin{acknowledgments}
 We thank Derek Meyers for useful comments. The work at Brookhaven National Laboratory was supported by the U.S. Department of Energy, Office of Basic Energy Sciences, Division of Materials Sciences and Engineering, under Contract No.\ DE-SC0012704. This research used resources 23-ID-1 beamline of the National Synchrotron Light Source II, a U.S. Department of Energy (DOE) Office of Science User Facility operated for the DOE Office of Science by Brookhaven National Laboratory under Contract No.\ DE-SC0012704. The sample pattern was performed at the Center for Functional Nanomaterials, which is a U.S. DOE Office of Science Facility, at Brookhaven National Laboratory under Contract No.\ DE-SC0012704. Work at Argonne National Laboratory was supported by the US Department of Energy, Office of Basic Energy Sciences, under Contract no. DE-AC0206CH11357.
\end{acknowledgments}

\subsection*{Author contributions}
S.B.W., M.P.M.D., and I.K.R.\ initiated and managed the project. X.M.C., C.M., Y.C., V.T., A.M.B., W.H., T.A., H.M., G.F., M.P.M.D., S.B.W., and I.K.R. prepared and performed the x-ray experiment. X.M.C, M.L.\ and G.D.G.\ prepared the sample. X.M.C., C.M., Y.C., V.T., J.M.T., M.P.M.D., S.B.W., and I.K.R. analyzed and interpreted the results. X.M.C., M.P.M.D., and I.K.R.\ wrote the manuscript.

\subsection*{Competing financial interests}
The authors declare no competing financial interests.

\end{document}